\documentclass{llncs}
\usepackage{listings}
\usepackage{amsmath,epsfig,epstopdf, amssymb}
\usepackage{algpseudocode}
\algnewcommand\algorithmicforeach{\textbf{for each}}
\algdef{S}[FOR]{ForEach}[1]{\algorithmicforeach\ #1\ \algorithmicdo}

\usepackage{wrapfig}
\usepackage{lipsum}
\usepackage{float}
\floatstyle{boxed}
\newfloat{algorithm}{htbp}{loa}
\floatname{algorithm}{Algorithm}
\newfloat{algorithm}{htbp}{loa}
\usepackage{wrapfig}

\usepackage{multirow}

\usepackage{etoolbox}\AtBeginEnvironment{algorithmic}{\small‌​}

\newtheorem{thm}{Theorem} 

\newtheorem{assumption}[thm]{Assumption}

\author{Omar Al-Bataineh$^{\star}$, Michael Fisher$^{ \dagger }$, and David Rosenblum$^{\star}$ }

\institute{$^{\star}$National University of Singapore, Singapore \\
			$^{ \dagger }$University of Liverpool, United Kingdom}
 
\title{Computing Maximal Expected Termination Time of Probabilistic Timed Automata}

\date{}

\begin{document}

\maketitle

\begin{abstract}

The paper addresses the problem of computing maximal expected time to
termination of probabilistic timed automata (PTA) models, under the
condition that the system will, eventually, terminate. This problem
can exhibit high computational complexity, in particular when the
automaton under analysis contains cycles that may be repeated very
often (due to very high probabilities, e.g. $p =0.999$). Such cycles
can degrade the performance of typical model checking algorithms, as
the likelihood of repeating the cycle converges to zero arbitrarily
slowly. We introduce an acceleration technique that can be applied to
improve the execution of such cycles by collapsing their iterations.
The acceleration process of a cyclic PTA consists of several formal
steps necessary to handle the cumulative timing and probability
information that result from successive executions of a cycle. The
advantages of acceleration are twofold. First, it helps to reduce the
computational complexity of the problem without adversely affecting
the outcome of the analysis. Second, it can bring the ``worst case
execution time'' problem of PTAs within the bounds of feasibility for
model checking techniques.  To our knowledge, this is the first work
that addresses the problem of accelerating execution of cycles that
exhibit both timing and probabilistic behavior.

\end{abstract}

\section{Introduction}
In this paper, we consider the problem of computing the ``expected
worst case execution time", or ``maximum expected termination time'',
for probabilistic timed automata (PTA). Given a probabilistic timed
automaton $\mathcal{P}$, with a start location $l_s$ and a final
location $l_f$, this problem aims to compute an upper bound on the
time needed to reach the final location $l_f$ from the start location
$l_s$. The problem is easy to solve in the case of acyclic PTA, but
successive executions of a cycle in a PTA model might yield a time
series whose total summation can potentially be unbounded. The problem
is interesting as cycles are common in the behavior of probabilistic
systems. It is important since, in modelling real, cyber-physical
systems, we often want to know not just ``how quickly'' but ``how
slowly'' a particular system might execute. In general, ``worst case
execution time'' (WCET) analysis is undecidable: it is undecidable to
determine whether or not an execution of a system will eventually
halt.  However, for PTA models one can often use model checking
 to analyse the system and compute the WCET.

The WCET problem for the case of \emph{non-probabilistic} timed systems with
cyclic behavior has been addressed in~\cite{BatainehTCS}, where a
model checking algorithm based on the zone-abstraction technique was
used allowing on-the-fly computation of WCET for timed automata models
and detection of the cases where WCET may be unbounded. For
\emph{probabilistic} timed systems, for example, the problem becomes
much harder, as any solution needs to handle both timed transitions
and probability distributions simultaneously. 

We present an efficient approach at computing the 
WCET of cyclic PTAs which attempts to avoid the
explicit repeated exploration of cycles encountered during model checking
(explicit-state exploration with clock zones computed to represent
the possible sets of values for a set of real-time clocks). 
This can be performed by detecting the
cycles, analyzing the periodic behavior of the cycles, collapsing the cycle
by computing the cummulative effect (in terms of contribution to WCET) of the
cycle, and then eliminating the cycle from the subsequent search. 
A key feature of the proposed WCET algorithm is that it can detect on-the-fly cycles in the input model and determine whether the detected cycle is a cycle with constant delays or a cycle with periodic delay  by examining only the characteristics of the reached fixed-points. 

The proposed algorithm is based, roughly, on extending the standard forward exploration of
the state space augmented with the acceleration of cycles encountered during the search, with some heuristics to optimize the computations. 
The primary case where the cycle collapsing presented in the algorithm would have
benefit is in systems where a cycle is taken with a high probability
potentially leading to numerous iterations before reaching some point of
escape. The proposed acceleration technique is an interesting addition to the collection of techniques for PTA analysis, where existing algorithms for PTAs \cite{KNPS03,Jovanovi2017}  are not optimized to check WCET.

\paragraph{\textbf{Related Work}.} The work in~\cite{KNPS03} studied 
the problem of computing expected costs or rewards in PTAs using
digital clocks, where they prove the equivalence of the continuous and
integer-time semantics w.r.t. expected rewards. The approach is
limited to finite-state models, and it is not clear how it performs in
presence of cycles that can be repeated with high probability.  The
authors have not proposed any acceleration technique to speed-up
the verification of WCET of cyclic PTAs. 

The work
in~\cite{Jovanovi2017} proposed a solution to the problem of computing
optimal expected reachability time in PTAs, relying on an
interpretation of the PTA as an uncountable-state Markov decision
process and employing a representation in terms of an extension of the
`simple' and `nice' functions of [4]. The optimal prices are computed
via a Bellman equation using value iteration. However, the authors did
not provide any details about the the complexity and efficiency of
their algorithm. It is also not clear how the algorithm behaves in
presence of complex cycles which can be repeated with high frequency.
Furthermore, the algorithm in~\cite{Jovanovi2017} does not employ any
form of acceleration technique to reduce the computational complexity
of the problem.

In~\cite{BatainehTCS}, the authors proposed a model checking algorithm
based on the zone abstraction for the problem of computing maximum
termination time of \emph{non-probabilistic} timed automata
(TA). However, for probabilistic timed systems the problem may be much
harder, as the solution needs to handle both timed transitions and
probability distributions. Moreover, the abstractions, optimisations,
and accelerations developed for the verification of WCET of
TAs~\cite{BatainehRF15,BatainehTCS} cannot be used to verify expected
WCET of PTAs, as cycles in PTAs exhibit both timing and probabilistic
behavior.

\section{Preliminaries}

In this section, for the sake of completeness, we recall the
definitions of probabilistic and timed probabilistic systems needed to
give semantics to probabilistic timed automata. We also recall
definitions of zone abstraction and the \emph{difference bound matrix}
data structure that is used to symbolically represent the state space
of probabilistic timed systems.

\subsection{Timed Probabilistic Systems}

A (discrete probability) distribution over a finite set $Q$ is a
function $\mu: Q \rightarrow [0, 1]$ such that $\sum_{q \in Q} \mu(q)
= 1$. For an uncountable set $Q^{'}$, let $\verb+Dist+(Q^{'})$ be the
set of distributions over finite subsets of $Q^{'}$.

\begin{definition} \textbf{(Probabilistic systems)}. \label{ProbSys} 
A probabilistic system \verb+PS+, is a tuple $(S, Steps, \mathcal{L})$
where $S$ is a set of states, $Steps \subseteq S \times
\verb+Dist+(S)$ is probabilistic transition relation, and
$\mathcal{L}: S \rightarrow 2^{AP}$ is a labelling function assigning
atomic propositions to states.

\end{definition}

\noindent A probabilistic transition $s \xrightarrow{\mu} s'$ is made
from a state s by nondeterministically selecting a distribution $\mu
\in \verb+Dist+(S)$ such that $(s, \mu) \in Steps$, and then making a
probabilistic choice of target state $s'$ according to $\mu$, such
that $\mu(s') > 0$. \\[1em]
We now consider the definition of timed probabilistic systems.

\begin{definition} \textbf{(Timed Probabilistic systems)}. A timed 
probabilistic system, \verb+TPS+, is a tuple $(S, Steps, \mathcal{L})$
where: $S$ and $\mathcal{L}$ are as in Definition \ref{ProbSys} and
$Steps \subseteq S \times \mathbb{R}\times \verb+Dist+(S)$ is a timed
probabilistic transition relation, such that, if $(s, t, \mu) \in
Steps$ and $t>0$, then $\mu$ is a distribution. 
 The component, $t$, of a tuple $(s, t, \mu)$ is called a
duration. 

\end{definition}

\subsection{PTA Models and Expected WCET Problem}

A probabilistic timed automaton
(PTA)~\cite{Alur1991,KwiatkowskaNSS02,Beauquier2003} models real-time
behaviour in the same fashion as a classical timed
automaton~\cite{Alur1994}, namely by using clocks. Clocks are
real-valued variables which increase at the same rate as time.  Let
$\mathcal{X}$ be the set of clock variables in a PTA $\mathcal{P}$.
We write $\mathcal{C}(\mathcal{X})$
to denote the set of clock constraints over $\mathcal{X}$, i.e., the
set of boolean combinations of atomic constraints of the form $x \sim
c$, where $\sim \in \{<, \leq, >, \geq\}$ and $c \in \mathbb{N}$.  We
note by $\mathcal{C}_{<\cdot}(\mathcal{X})$ the restriction of
$\mathcal{C}(\mathcal{X})$ to positive boolean combinations only
containing constraints of the form $x \leq c$ or $x < c$.

\begin{definition} (\textbf{PTA syntax}). A probabilistic timed 
automaton (PTA) is defined by a tuple $\mathcal{P} = (L, \ell_0, L_f,
\mathcal{X}, Act, inv, E, \mathcal{L})$ where
\begin{itemize}

\item $L$ is a finite set of locations and $\ell_0 \in L$ is an initial location;

\item $L_F \subseteq L$ is a finite set of final (halting) locations;

\item $\mathcal{X}$ is a finite set of clocks;

\item $Act$ is a finite set of actions;

\item $inv: L \rightarrow \mathcal{C}_{<\cdot}(\mathcal{X})$ is an invariant condition;

\item $E \subseteq L \times Act \times \mathcal{C}(\mathcal{X}) \times Dist(2^{\mathcal{X}} \times L) \times L$ is a finite set of probabilistic edges;


\item $\mathcal{L}: L \rightarrow 2^{AP}$ is a labelling function mapping each location to a set of atomic propositions. 

\end{itemize}

\end{definition}

\begin{definition} (\textbf{PTA Semantics}). 
Let $\mathcal{P} = (L, \ell_0, L_f, \mathcal{X}, Act, inv, E, \mathcal{L})$ be a PTA. The semantic of $\mathcal{P}$ is defined as the (infinite-state) timed probabilistic system $TPS_{\mathcal{P}} = (S, Step, \mathcal{L^{'}})$ where $S \subseteq L \times \mathbb{R}^{X}$ such that $(\ell, v) \in S$ if, and only if, $v \models inv(\ell)$ and $(\ell, v), t, \mu \in Steps$ if and only if the following conditions hold

\begin{itemize}

\item Time transitions: $t\geq 0$, $\mu = \mu(\ell, v+t)$ and $v +
  t^{'} \models inv(\ell)$ for $0 \leq t^{'} \leq t$

\item Discrete transitions: $t=0$ and there exists $(\ell, a, g, d, \ell^{'}) \in E$ such that $v \models g$ and for any $(\ell^{'}, v^{'}) \in S: \mu(\ell^{'}, v^{'}) = \sum_{X \subseteq \mathcal{X} \land v^{'} =v[X:=0]}d(X, \ell^{'}) $

\end{itemize}

\end{definition}

\noindent A state of a PTA is a pair $(\ell, v) \in L \times
\mathcal{R}_{\geq 0}^{\mathcal{X}}$ such that $v \models
inv(\ell)$. In any state $(\ell, v)$, either a certain amount of time $t
\in \mathcal{R}_{\geq 0}$ elapses, or an action $a \in Act$ is
performed. If time elapses, then the choice of $t$ requires that the
invariant $inv(\ell)$ remains continuously satisfied while time
passes.  
We write $(\ell, v) \xrightarrow{t, e, p} (\ell^{'}, v^{'})$
if from state $(\ell, v + t)$ and assuming probabilistic edge $e$ is selected,
the next state is $(\ell^{'}, v^{'})$ with probability $p$.
Throughout this paper, we use the following
notations: $weight(e)$ to refer to the probability weight of an edge
$e$, $src(e)$ to refer to the source control location of edge $e$, and
$out(src(e))$ to refer to the set of outgoing edges of the location $src(e)$.
For example, if $e =(\ell, a, g, d, \ell^{'})$ then 
$src(e) = \ell$ and $weight(e) = d (X, \ell^{'}) = p$, where $p \in (0, 1]$ and $X \subseteq \mathcal{X}$. However, in this paper, we make the following assumptions on the PTAs we consider.

\begin{assumption} \label{Assumptions} For any PTA $\mathcal{P}$ we have:

\begin{enumerate}

\item  all states in $\mathcal{P}$ behave purely probabilistic
(i.e. there is no non-determinism between edges of $\mathcal{P}$);

\item  every probabilistic edge in $\mathcal{P}$ is associated with a weight from (0, 1];

\item $\mathcal{P}$ is a flat automaton, where each location 
in $\mathcal{P}$ is part of at most one cycle;

\item $\mathcal{P}$ is structurally non-zeno;

\item $\mathcal{P}$ is well-formed (i.e. all transitions in $\mathcal{P}$ lead to valid states);

\item all invariants of $\mathcal{P}$ are bounded;

\item  halting states of $\mathcal{P}$ are time-lock states;

\item all invariants and enabling conditions of $\mathcal{P}$ are convex;

\end{enumerate}

\end{assumption}

\noindent It is interesting to note that in PTAs, edges do not result
in the reset of a fixed set of clocks leading to a fixed location, but
rather yield a distribution $ d \in Dist(2^{\mathcal{X}} \times
L)$ over resets and locations.  Hence, a run of a PTA can be split
into several parallel subruns whenever the nodes of PTA have
probabilistic choices.  It thus may seem natural to define a run of
PTA as a tree (i.e. set of branches) whose nodes are labeled by
configurations of the automaton.  To simplify definition of WCET of
 PTAs, we will consider symbolic runs, that is, a special
sets of runs in PTAs in which the time delay the automaton can spend
at a control location $\ell_i$ is represented by an interval $T_i$ of
the form $[T^{\min}_{i}, T^{\max}_{i}]$. We can then define WCET as follows.

\begin{definition} \label {PurelyPTAs} (\textbf{WCET of PTAs}). 
Let $\mathcal{P}$ be a single-run PTA with a symbolic run $r$. 
Suppose that $r$ can be split into $r_0,..., r_{k-1}$  symbolic subruns, where each subrun $r_i$
has the form $\ell_{0} \xrightarrow{T_0, e_0, p_0} \ell_{1} \xrightarrow{T_1, e_1, p_1} .. \xrightarrow{T_{n-1}, e_{n-1}, p_{n-1}} \ell_{f}$. 
Then maximum delay of $r_i$ can be computed as follows
$$
Maxdelay(r_i) = \sum_{a = 0}^{n-1} (\prod_{b=0}^{a} (p_{b})) * T^{\max}_{a}.
$$

Hence, WCET of $\mathcal{P}$ can be computed as follows
$$
\mathbb{WCET}  (\mathcal{P}) =  \sum_{ i = 0}^{k-1} Maxdelay (r_i). 
$$
\end{definition}

\begin{definition} (\textbf{Termination of PTAs}).
We say that a PTA $\mathcal{P}$  with single-run  $r$ terminates if 
every subrun of $r$ reaches a halting state. 
\end{definition}

\begin{wrapfigure}{R}{0.5\textwidth} 
\vspace{-20pt}
     \centering
    \includegraphics[width=0.48\textwidth] {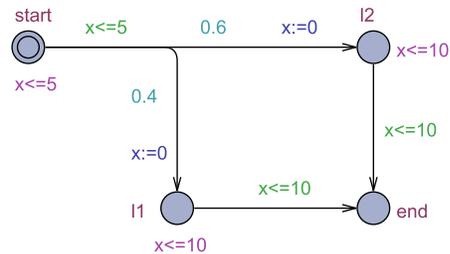}
    \caption{$\mathcal{P}_1$: A PTA with two subruns}
    \label{fig:deterPTA}
     \vspace{-20pt}
  \vspace{1pt}
\end{wrapfigure}

\paragraph{Example 1.} To demonstrate how one can compute WCET of PTAs,
let us consider the PTA given in Fig. \ref{fig:deterPTA}.
Note that the given PTA consists of a single run 
which can be split into two subruns $r_1$ and $r_2$,
where $r_1$ visits locations \verb+start+,  \verb+l1+ and \verb+end+
while subrun $r_2$ visits locations \verb+start+,  \verb+l2+ and \verb+end+.
The location \verb+end+ represents the halting location of the automaton.
Let us denote the edge from \verb+start+ to  \verb+l1+ by $e_0$
 and the edge from \verb+start+ to  \verb+l2+ by $e_1$.
 Then $weight(e_0) = 0.4$ and  $weight(e_1) = 0.6$,
 while the other edges have probability weight of one.
The WCET of this automaton can be obtained by taking the sum
of the delays of the two subruns, while maximizing the 
time delay the automaton can spend at each visited location.
That is, $Maxdelay(r_1) = 5 * 0.4 + 0.4 *10 = 6$
and $Maxdelay(r_2) = 5 * 0.6 + 0.6 *10 = 9$.
Hence, $\mathbb{WCET}  (\mathcal{P}_1) = Maxdelay(r_1) + Maxdelay(r_2) = 15$.

 Our goal here is to develop an efficient solution for WCET
 of cyclic PTAs by accelerating the execution of cycles that
can be taken with high probability.  Such classes of cycles
can degrade the performance of model checking algorithms, since the
probability to repeat the cycle converges to zero arbitrarily  slowly.

\begin{wrapfigure}{r}{0.5\textwidth} 
\vspace{-20pt}
     \centering     \centering
    \includegraphics[width=0.5\textwidth]{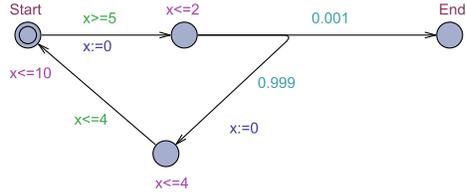}
    \caption{$\mathcal{P}_2$: A cyclic PTA which can be taken with high probability}
    \label{fig:slowRate}
    \vspace{-20pt}
  \vspace{1pt}
\end{wrapfigure}

\paragraph{Example 2.} The automaton in Fig.~\ref{fig:slowRate} contains a cycle
that can be repeated with very high probability, where the likelihood
to repeat the cycle gradually decreases. 
The first time we reach the choice point in Fig.~\ref{fig:slowRate},
the probability of cycling is $0.999$ while the probability of moving
to the `End' state is $0.001$. If we take the cycle then the next time
we reach the choice point we will effectively have a lower probability
of again taking the cycle `choice'. Effectively, the probability here
is $0.999\times 0.999$. And so on. In this way, the likelihood of
staying within the cycle monotonically decreases and eventually reaches
zero (or close enough to be considered as zero).

\subsection{The Zone Abstraction and Difference Bound Matrices}
The state space of dense-time models can, in general, be infinite
(uncountable) and therefore can not be directly model
checked. However, researchers in real-time model checking devised an
efficient representation of the state space of a TA based on
\emph{zone-graphs}~\cite{Henzinger92,Yannakakis1997}. In a zone graph,
zones denote symbolic states. In practice, this provides a more
compact representation of the state-space of a given TA model.

A zone is a pair $(l, \varphi)$, where $l$ is a location of a PTA $\mathcal{P}$ and $\varphi$ is a clock zone. The clock zone $succ(\varphi, e)$ will denote the set of clock valuations $v^{'}$ such  that for some $v \in \varphi$ the state $(l^{'}, v^{'})$ can be reached from the state $(l, v)$ by letting time elapse and by executing the transition $e$.
The pair $(l^{'}, succ(\varphi,e))$ will represent the set of successors of $(l, \varphi)$
under the transition $e$.  Note that the assignment of the values of the clocks in the initial location of $\mathcal{P}$ is easily expressed as a clock zone since $v(x) = 0$ for every clock $x \in X$. Note also that every constraint used in the invariant of an automaton location or in the guard of a transition is a clock zone. Therefore, clock zones can be used for various state reachability analysis algorithms for (probabilistic) timed automata.  

\emph{Difference bound matrices} (DBMs)~\cite{Dill1990} are the data
structures most commonly used for representing the state spaces of
(probabilistic) timed automata. A DBM is a two-dimensional matrix that
records the difference between upper bounds of clock pairs up to a
certain constant. Recall that a clock constraint over the set of
clocks $X$ is a conjunction of atomic constraints of the form $x \sim
m$ and $x-y \sim n$ where $x, y \in X$, $\sim \in \{ \leq, <, =, \geq,
>\}$, and $m, n$ are integers. In order to provide a unified form for
clock constraints in a DBM we introduce a reference clock $x_{0} \in
X$ with the constant value 0 that is not used in any guards or
invariants. The matrix is indexed by the clocks in $X$ together with
the special clock $x_{0}$. The element $D_{i,j}$ in matrix $D$ is of
the form $(n, \prec)$ where $x_{i}, x_{j} \in X$, $n$ represents the
difference between them, and $ \prec \in \{ \leq, <\}$.  Each row in
the matrix represents the bound difference between the value of the
clock $x_{i}$ and all the other clocks in the zone, thus a zone can be
represented by at most $|X|^2$ atomic constraints. This implies that
each pair of variables $(x_{i},x_{j})$ ($i \neq j$) will be
represented by two atomic constraints $(d_{i,j}, \prec)$ and
$(d_{j,i}, \prec)$.

\section{Accelerating Execution of Probabilistic Timed Cycles} 
\label{sec: accelerations}
In this section we discuss some acceleration techniques that can be
used to improve the execution of cycles that may be repeated a high
number of times. Let us denote the series of maximal expected delays that results from successive
executions of a cycle $\pi$ in a PTA $\mathcal{P}$
by $\mathcal{S}_{\pi}$ and $n$ be a cycle counter.  
We then use the notations $\mathcal{S}_{\pi} \xrightarrow{p} 0$ to
denote that the probability of taking the cycle moves to zero, and
$\mathcal{S}_{\pi} \xrightarrow{a.s} \zeta$, where $\zeta < \infty$,
to denote that the series converges \emph{almost surely}.
However, for any reachable cycle $\pi$ in the PTAs we consider, the
series $\mathcal{S}_{\pi}$ converges to zero probability (i.e. the
cycle will not be taken forever) and converges with probability one,
as discussed in Theorem~\ref{AlmostSurelyConverg}. (Note that, as the
effective probability of remaining in a cycle reduces every time we
take the cycle, we often view the probability at the branch point as
reducing in this way.)

\begin{theorem} \label{AlmostSurelyConverg}
Let $\pi$ be a cycle in a PTA model that satisfies Assumption \ref{Assumptions}.
Then (1) $\mathcal{S}_{\pi} \xrightarrow{p} 0$ as $n \rightarrow \infty$ 
and (2) $\mathcal{S}_{\pi} \xrightarrow{a.s} \zeta$ as $n \rightarrow \infty$, where $\zeta < \infty$.
\end{theorem}

\begin{figure}
    \begin{minipage}{.5\textwidth}
     \centering
    \includegraphics[width= 2.2in]{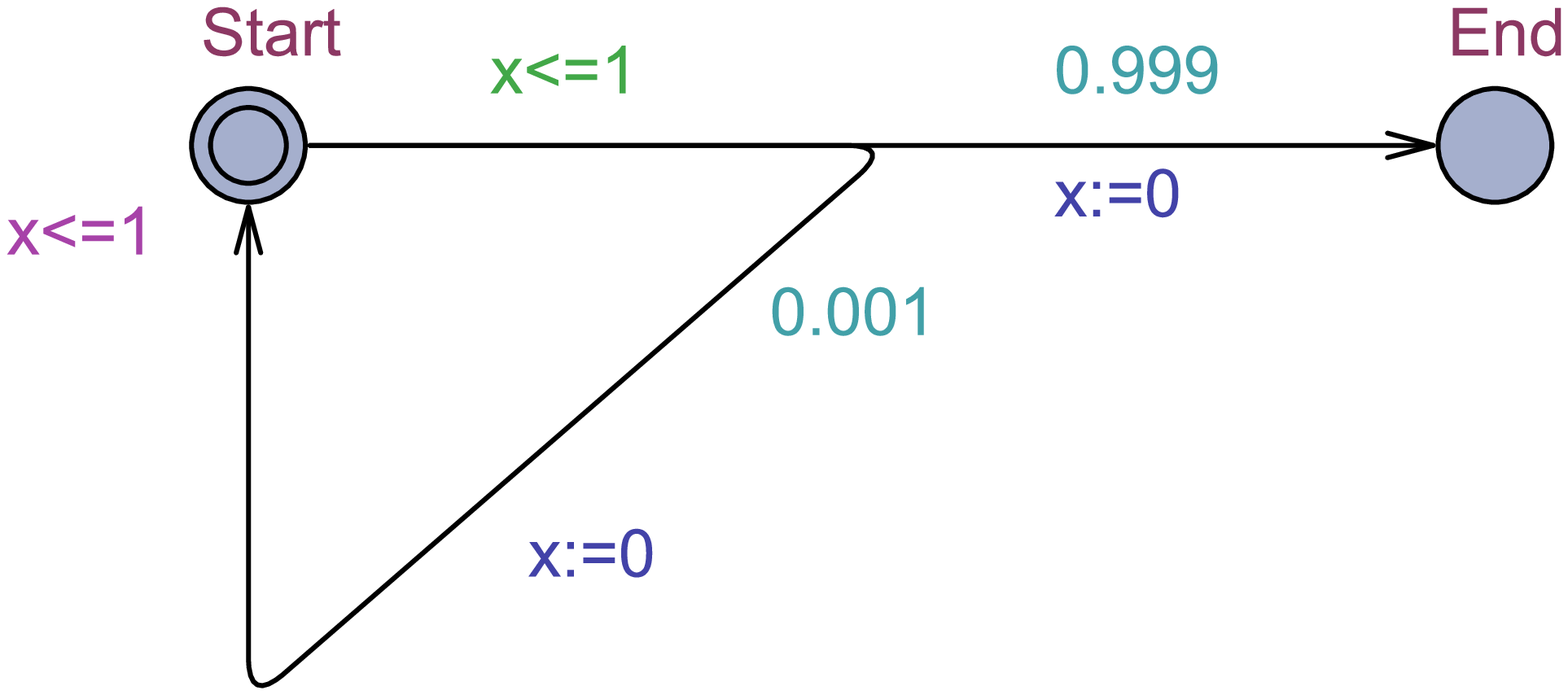}
    \caption {$\mathcal{S}_{\pi}$ converges sufficiently fast}
    \label{fig:fast}
    \end{minipage}
    \begin{minipage}{.5\textwidth}
        \centering
    \includegraphics[width= 2.2 in]{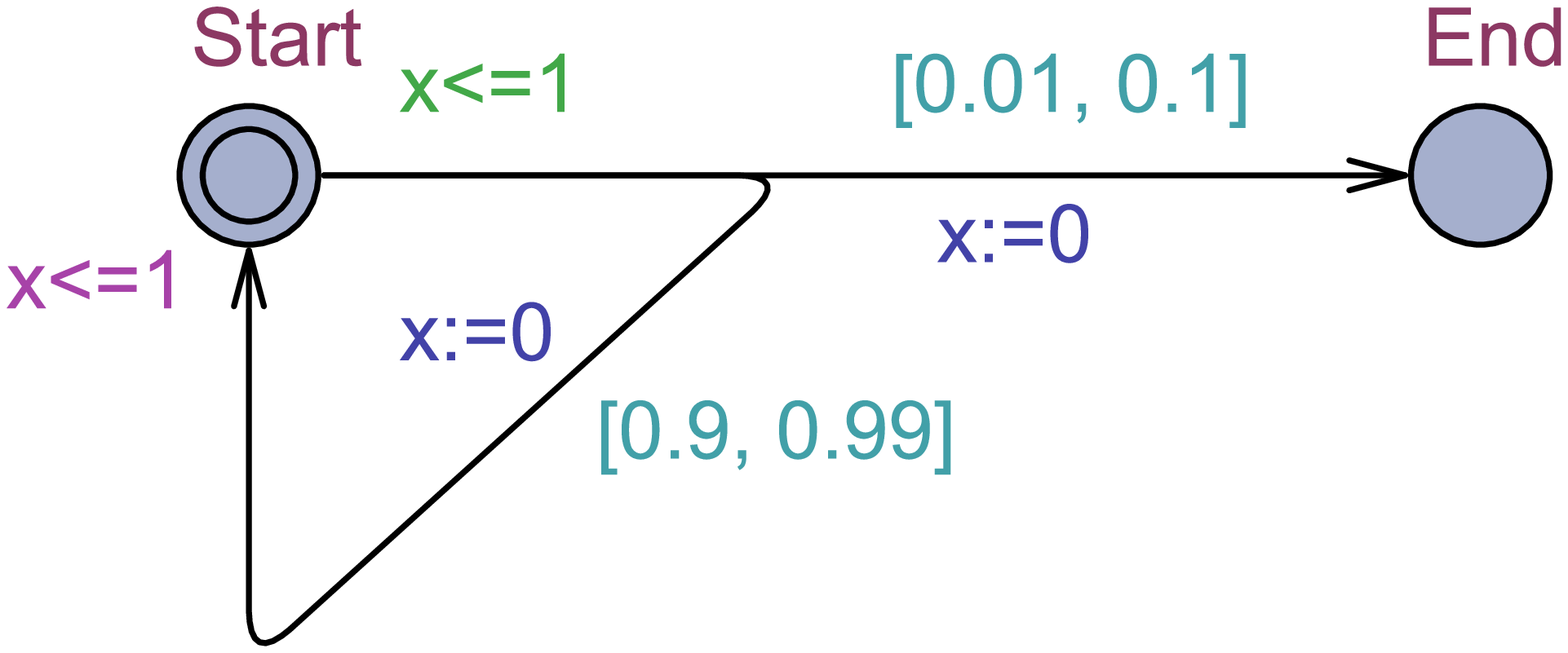}
    \caption{$\mathcal{S}_{\pi}$ converges arbitrary slow}
    \label{fig:slow}
    \end{minipage}%
 \end{figure}

Note that the series $\mathcal{S}_{\pi}$ may converge to zero
probability sufficiently fast or arbitrary slow depending on the
probability weights of the edges of the cycle.  Suppose that we use an
approximation bound $\Delta = 10 ^{-6}$ to represent ``close enough to
zero'' in probability when executing cycles in PTAs.  So that once the
probability that results from successive executions of a cycle becomes
smaller than the bound $\Delta$, the cycle will no longer be repeated.
It is easy to see then that the cycle in Fig. \ref{fig:fast} will be repeated only four 
iterations (as it becomes ``close enough to zero'' quite quickly) where the series $\mathcal{S}_{\pi} = (10^{-2} + 10^{-4} + 10^{-6} + 10^{-8})$,
while the cycle in Fig. \ref{fig:slow} will be repeated around $13808$ iterations
where the series $\mathcal{S}_{\pi}= (0.999 + 0.998001 + 0.997002 + 0.996005 + ... + 9 \times 10 ^{-8})$. 
 We now discuss two forms of cycles that may be encountered when analyzing a cyclic PTA model:
cycles with constant delays and cycles with periodic delays.

\begin{definition} (\textbf{Cycles with constant delays}). \label{ConstantDelays}
Let $\pi$ be a cycle in a PTA  model $\mathcal{P}$ and  $\verb+delay+(\pi, i)$ be a function that computes the summation of delays of $\pi$ at some arbitrary iteration $i \geq 1$. We say that $\pi$ is a cycle with constant delays if for any two distinct iterations $i, j$ we have $\verb+delay+(\pi, i) = \verb+delay+(\pi, j)$.
\end{definition}

\begin{definition} (\textbf{Cycles with periodic delays}). \label{PeriodicDelay}
Let $\pi$ be a cycle in a PTA  model $\mathcal{P}$. We say that $\pi$ is a cycle with periodic delays if the delays of $\pi$ are repeated every $k$ iterations, where $k > 1$.
That is, $\verb+delay+(\pi, i) = \verb+delay+(\pi, i + k)$.

\end{definition}

\noindent We now describe the basic formal steps that can be followed
to accelerate the execution of a cycle $\pi$ in a PTA model
$\mathcal{P}$.

\begin{enumerate}

\item Synthesize a delay formula, $\phi_{\pi}$, for the detected cycle
  $\pi$ that can be used to compute the cumulative delay introduced by
  successive executions of $\pi$.  A delay formula for $\pi$ can be
  synthesized once a fixed-point of $\pi$ is reached.
 
\item Find the value of the loop counter $n$ at 
which the probability to repeat the cycle converges to zero.
Recall that for PTAs we consider the probability to repeat cycles
decreases monotonically as the iteration number increases. 

\item Compute the total expected delay of the cycle $\pi$ using $\phi_{\pi}$.

\item Compute the clock zone that results from collapsing the cycle's iterations.

\item Update the probability weights of the automaton edges that have been affected by the acceleration process. 

\item  Restart the corresponding constructed Markov chain of $\mathcal{P}$.

\end{enumerate}
We first discuss how one can synthesize a formula for computing
expected delay of cycles with constant delays and cycles with periodic
delays.

\begin{definition} (\textbf{Synthesizing formulae for cycles with constant delays}). \label{DefnFormula}  Let $\mathcal{P} = (L, \ell_0, L_f, \mathcal{X}, Act, inv, E, \mathcal{L})$ be a PTA and  $(e_0,...,e_{m-1}) \in E_{\pi}$ be the sequence of edges of a reachable cycle $\pi$ in $\mathcal{P} $ whose delay intervals between iterations are constant.  Let $T_0^{\max},...,T_{m-1}^{\max}$ be the maximum delay bounds that can elapse at the cycle's locations $src(e_0),..., src(e_{m-1})$. The cumulative delays that result from successive executions of $\pi$ can be computed as follows

$$ \phi_{\pi}^{c} = \sum_{a = 0}^{n} \sum_{b =0}^{m-1} \mathcal{I} *
  (\prod_{c = 0}^{m-1} weight(e_c))^{a} * (\prod_{d = 0}^{b-1}
  weight(e_d)) * T_b^{\max}
$$

\end{definition}
where $\mathcal{I}$ represents the initial probability value at which
the cycle $\pi$ has been reached during the analysis. 
The first summation operator in the formula is used to iterate through the cycle
until the probability to repeat the cycle effectively converges to
zero, while the second summation operator is used to iterate through
the control locations of the cycle at each iteration.  Since the
control locations of the cycle can be reached with different
probability values at each different iteration, the expected delays
that result from visiting these locations can vary between iterations.
However, the formula in Definition \ref{DefnFormula} can be simplified
further as $\sigma = \prod_{c = 0}^{m-1} weight(e_c)$ is constant.
This yields the following formula
$$\phi_{\pi}^{c} = \sum_{a = 0}^{n} \sum_{b =0}^{m-1} \mathcal{I}  * \sigma^{a} * (\prod_{c = 0}^{b-1} weight(e_c)) * T_b^{\max}.$$ 

It remains to discuss how to compute the value of $n$ (i.e. the number of times the cycle can be repeated).  To find the value of $n$ we need to solve the simple exponential formula
$
\sigma^{n} = \dfrac{\Delta}{\mathcal{I}}$.
However, by taking the natural logarithmic of both sides, then $n$ can be computed as follows
$$
n =  (\dfrac{ln(\Delta)}{ln(\sigma * \mathcal{I})}).
$$

\begin{definition} (\textbf{Synthesizing formulae for cycles with periodic delays}). \label{PeriodicFormula}  Let $\mathcal{P} = (L, \ell_0, L_f, \mathcal{X}, Act, inv, E, \mathcal{L})$ be a PTA and  $(e_0,...,e_{m-1}) \in E_{\pi}$ be the sequence of edges of a reachable cycle $\pi$ in $\mathcal{P}$. Suppose that $\pi$ is a cycle with periodic delays so that the delays are repeated every $k$ iterations.   The cumulative delays that result from successive executions of $\pi$ can be computed as follows
$$
\phi_{\pi}^{p} = \sum_{a = 0, a+ k }^{n} \sum_{b = 1}^{k}   \sum_{c =0}^{m-1} \mathcal{I}  *(\prod_{d = 0}^{m-1} weight(e_d))^{a} * (\prod_{e = 0}^{b-1} weight(e_e)) * T_{(b, c)}^{\max}
$$
where $k$ represents the rate (i.e. number of iterations) at which delays of $\pi$ are repeated, and $T_{(b, c)}^{\max}$ is the maximum delay that $\mathcal{P}$ can spend at location $src(e_c)$ at iteration $b$ where $b \in \{1,..k\}$ and $c \in \{0,..,m-1\}$. However, since the cycle contains periodic delays then every $k$ iterations the counter $b$ needs to be reset. Similar to cycles with constant delays, the given formula can be simplified to

$$
\phi_{\pi}^{p} = \sum_{a = 0, a + k }^{n} \sum_{b =0}^{k-1}   \sum_{c =0}^{m-1} \mathcal{I}  * \sigma^{a} * (\prod_{d = 0}^{b-1} weight(e_d)) * T_{(b, c)}^{\max}.
$$
\end{definition}

\noindent The next step in the process is to compute the accelerated
clock zone that results from collapsing iterations of the cycle.
Recall that zones provide a representation of sets of clock
interpretations as constraints on (lower and upper) bounds on
individual clocks and clock differences. Let $k$ be the iteration
number at which the delay of the cycle becomes constant and $n$ be the
value of the cycle counter at which the probability to repeat the
cycle effectively converges to zero. We can compute the clock zone
that results from accelerating such a cycle as follows.

\begin{itemize}
 
 \item \emph{Updating lower/upper bounds of the automaton clocks.}
   Updating the automaton clocks during acceleration is an easy task
   as the delays of the cycle are constant between iterations. Hence,
   the lower and upper bound of a clock $z$ can be updated as follows.
$$
D_{0, z}^{n} = (D_{0, z}^{k} + (n-k) *  (D_{0, z}^{k} - D_{0, z}^{k-1}))
$$
$$
 D^{n}_{z, 0} = (D_{z, 0}^{k} + (n-k) *  (D_{z, 0}^{k} - D_{z, 0}^{k-1} ))
$$

\item \emph{Updating diagonal constraints of the automaton clocks.}
  Updating this set of constraints is also a straightforward task. Let
  $z_1$ and $z_2$ be two clocks in the automaton being
  accelerated. Then the diagonal constraints involving $z_1$ and $z_2$
  can be updated as follows.
$$
D_{z_1, z_2}^{n} = D_{z_1, z_2}^{k} + (D^{k}_{z_1, 0} - D^{k}_{0, z_2}) * (n-k)
$$
$$
D_{z_2, z_1}^{n} = D_{z_2, z_1}^{k} + (D^{k}_{z_2, 0} - D^{k}_{0, z_1}) * (n-k)
$$

\end{itemize}
We now turn to discuss how to compute the accelerated clock zone that
results from collapsing the iterations of cycles with periodic
delays. For this class of cycles, the lower and upper bounds of the
automaton clocks can be updated as follows, where the variable $k$
used in the formulae to represent the rate (number of iterations) at
which delays are repeated.

$$
D_{0, z}^{n} = ((D_{0, z}^{k} - D_{0, z}^{0}) * \lfloor n / k \rfloor) + \sum_{i=1}^{(n \% k)} (D_{0, z}^{i} - D_{0, z}^{i-1})
$$ 
$$
D_{z, 0}^{n} = ((D_{z, 0}^{k} - D_{z, 0}^{0}) * \lfloor n / k \rfloor) + \sum_{i=1}^{(n \% k)} (D_{z, 0}^{i} - D_{z, 0}^{i-1})
$$ 
The diagonal constraints of the automaton clocks can be updated as follows.
$$
D_{z_1, z_2}^{n} = ((D_{z_1, z_2}^{k} - D_{z_1, z_2}^{0}) * \lfloor n / k \rfloor) + \sum_{i=1}^{(n \% k)} (D_{z_1, z_2}^{i} - D_{z_1, z_2}^{i-1})
$$
$$
D_{z_2, z_1}^{n} = ((D_{z_2, z_1}^{k} - D_{z_2, z_1}^{0}) * \lfloor n / k \rfloor) + \sum_{i=1}^{(n \% k)} (D_{z_2, z_1}^{i} - D_{z_2, z_1}^{i-1})
$$
The next important step of the acceleration process is to update the
probability weights of the edges of the automaton that have been
affected by the acceleration.
Note that, after acceleration, the probability weights of some edges
of the cycle will have decreased (e.g. will be set to zero) and hence
the probability weights of some other edges of the automaton need to
be updated (increased) in order to maintain the overall probability
distribution at states. This step can also be performed according to
the update rules given in Definition~\ref{updateRule}.

\begin{definition} (\textbf{Probability update rules after acceleration}). \label{updateRule}
Let $\mathcal{P} = (L, \ell_0, L_f, \mathcal{X}, Act, inv, E, \mathcal{L})$ be a PTA and  $(e_0, e_1,...,e_{m-1}) \in E_{\pi}$ be the sequence of edges of a reachable cycle $\pi$ in $\mathcal{P} $. Then after accelerating the execution of $\pi$ the probability weights of some edges in $\mathcal{P}$ will be updated as follows

\begin{enumerate}

\item Let $E_{out}$ be the set of edges in the set $out(src(e_i)) \setminus e_i$, where $e_i \in E_{\pi}$. Then for each edge $e_j \in E_{out}$, such that $0 < weight(e_j) < 1$, update the probability weight of $e_j$ as follows
$$
\overline{weight(e_j)} = weight(e_j) + \frac{weight(e_j) * weight(e_i)} {\sum_{e_k \in E_{out}} weight(e_k)} 
$$
where $weight(e_j)$ represents the probability weight of the edge $e_j$ in the prior distribution (before acceleration) and $\overline{weight(e_j)}$ represents the probability weight of the edge $e_j$ in the new distribution (after acceleration).

\item For each edge $e_i \in E_{\pi}$ whose $weight(e_i) < 1$ set $weight(e_i)$ to zero.

\end{enumerate}

\end{definition}

\noindent 
The last step of the process involves restarting the
Markov chain of the model $\mathcal{P}$ by setting the initial
probability of the system to one. The new initial state of the model
will be chosen according to available probabilistic choices.

\subsection{Effectiveness of Acceleration}

The effectiveness of the proposed acceleration (the possible reduction
on the size of the generated zone graph) depends on four factors:
(a) the value of $\sigma$ (the rate at which the probability to repeat the cycle is decreasing),
(b) the length of the cycle being accelerated, 
(c) the approximation bound $\Delta$ used to represent ``convergence to zero'',
and (d) the size of the states of the model (the number of the clocks in the model
as this can affect the size of the generated DBMs).

\begin{theorem} (\textbf{Effectiveness of acceleration}).
Let $\mathcal{P}$ be a PTA that satisfies Assumption \ref{Assumptions}
and $\pi$ be a reachable cycle in $\mathcal{P}$. Then the proposed 
acceleration can reduce the size of the generated zone graph
of $\mathcal{P}$  by $((n - k) \times length(\pi))$ states, where $n$ represents
the number of times the cycle can be repeated, $k$ represents 
the iteration number at which a fixed-point of $\pi$ can be reached, 
and $length(\pi)$ represents the number of transitions of $\pi$. 
\end{theorem}

\noindent Let us denote the zone graph that results from model
checking the non-accelerated cyclic PTA automaton $\mathcal{P}$ in
which all system states are explored by $\mathcal{Z}(\mathcal{P})$,
and the graph that results from model checking the accelerated version
of $\mathcal{P}$ where cycles iterations are collapsed by
$\mathcal{Z}(\mathcal{P^{\mathfrak{a}}})$.  Suppose that branches or
subruns of $\mathcal{P}$ contain $m$-cycles $\{\pi_{1}, ..., \pi_{m}
\}$.  Then the reduction gained ($RG$) from accelerating the
executions of cycles in $\mathcal{P}$ can be measured as follows
$$
RG = (|\mathcal{Z}(\mathcal{P})| - |\mathcal{Z}(\mathcal{P^{\mathfrak{a}}}|) 
   = \sum_{i =1}^{m} ((n_i - k_i) \times length(\pi_i)).
$$
%
The reader can easily construct an example where the series
$\mathcal{S}_{\pi}$ (the series that results from successive
executions of $\pi$) converges almost surely while the series
converges to zero probability arbitrarily  slowly. 





\section{A Zone-based Algorithm for WCET of  Cyclic PTAs} \label{sec:Algorithm}
In this section, we describe a zone-based algorithm that can be used
to compute the expected WCET of cyclic PTAs.  Each node in the
computed zone graph of the given PTA model has the form $(\ell, Z,
\alpha, sts, cnt)$ where the variable $sts$ (which is assigned to each
state) is used to detect whether there exists a cycle on locations in
the behavior of the automaton. The variable $sts$ can take values from
the set $\{0, 1, 2 \}$. When it is 0 it means that the location has
not been visited before, when it is 1 it means the location has been
visited before but not fully explored, and when it is 2 it means that
everything reachable from that location has been explored.  We assume
that the reader is familiar with the classical DFS algorithm with the
labeling process of nodes to unvisited (0), being explored (1), and
finished (2) and hence we omit these details.  The variable $\alpha$
maintains the probability value at which the state has been reached.
The variable $cnt$ is used to keep track of the iteration number of a
detected cycle, where $cnt$ is incremented every time a full iteration
of the cycle is completed and reset once the cycle is skipped.  By
examining the value of the variable $cnt$ when a fixed-point of a
cycle is reached, we can then distinguish between different forms of
cycles.

\begin{definition} \label{counter} (\textbf{Detecting cycles with constant/periodic delays}.)
Let $\pi$ be a reachable cycle in a PTA model $\mathcal{P}$.
Suppose that during the analysis of $\pi$ the two states $s$ and $s'$ have been reached
where $(s.\ell = s'.\ell \land  (s.Z \setminus inact = s'.Z\setminus inact))$ (i.e. a fixed-point has been reached w.r.t. the active clocks of the cycle).
Suppose further that $s.cnt < s'.cnt$ so that the state $s'$ has been reached
in an iteration that is greater than state $s$.  We can then determine the class of the cycle $\pi$
by examining the characteristics of the reached fixed-point as follows 

\begin{enumerate}

\item We say that $\pi$ is a cycle with constant delays or a cycle whose delays become constant after some iterations if the following condition holds
$$
(s.\ell = s'.\ell \land  (s.Z \setminus inact = s'.Z\setminus inact) \land (s'.cnt \leq 3 \lor  (s'.cnt - s.cnt) = 1))
$$

\item We say that $\pi$ is a cycle with periodic delays if the following condition holds
$$
(s.\ell = s'.\ell \land  (s.Z \setminus inact = s'.Z\setminus inact) \land s'.cnt > 3 \land  (s'.cnt - s.cnt) > 1)
$$

\end{enumerate}
\end{definition}

\noindent It is interesting to note that the set of clock zones that
result from the first iteration of a cycle can be arbitrary zones as
the initial zone at which the cycle is reached has not been obtained
from the cycle's internal computations. Hence, if a fixed-point of a
cycle is reached within the first three iterations, or within any two
consecutive iterations of the cycle, then we know that the cycle must
have constant delays. Otherwise, the cycle will have periodic
delays. Note that the tests described in Definition~\ref{counter} can
detect all forms of cycles with constant or periodic delays,
regardless of their underlying syntactic structures.

\begin{algorithm} [h!]
\begin{algorithmic} [1]
\caption{An algorithm for computing WCET of deterministic PTAs}
\label{alg:EfficientWCETWithAccel}
\State \textbf{Input}: $(\mathcal{P}$)
\State \textbf{Output}:  $\textbf{double} ~ \verb+WCET+ := 0$
\State \textbf{double} $\alpha := 1, prob: =1, \Delta:= 10^{-6}$
\State \textbf{int} $sts:=0, cnt  :=0 $ 
\State \textbf{clock} $\verb+CLK+$
\State WAIT := $\{(l_{0}, Z_{0}, \alpha, sts,  cnt) \}$,  PASSED := $\emptyset$ 
\While{WAIT $\neq ~ \emptyset$}
\State select $s$ from WAIT 
\State \textbf{add} $s$ to PASSED 
\ForEach{$e \in out(s.\ell)$}
\State $prob := (s.\alpha *  weight(e))$ 
\State $s' := succ(s.Z, e)$
\If {$s'.\ell = s^{''}.\ell \land s^{''}.sts = 1 \land s'.Z = s^{''}.Z$ for any $s^{''} \in$ PASSED}
\State $L_{\pi}$ := $ComputeLocationsofCycle()$
\State $\phi_{\pi} := SynthDelayFormula(L_{\pi})$
\If {$s'.cnt \leq 3 ~ \lor ((s'.cnt - s^{''}.cnt) = 1) $} 
\ForEach {$\ell \in L_{\pi}$ \textbf{such that} $out(\ell) > 1}$
\State \textbf{add} $(\ell, AccelConstCycle(L_{\pi}))$ \textbf{to} WAIT
\EndFor
\ElsIf {$s'.cnt > 3 ~ \land ((s'.cnt - s^{''}.cnt) > 1)$} 
\ForEach {$\ell \in L_{\pi}$ \textbf{such that} $out(\ell) > 1}$
\State \textbf{add} $(\ell, AccelPeriodCycle(L_{\pi}))$ \textbf{to} WAIT
\EndFor
\EndIf
\State $s'.cnt:=0$ 
\ElsIf {$s'.\ell = s^{''}.\ell \land s^{''}.sts = 1 \land s'.\dot{Z} \neq s^{''}.\dot{Z} \land   prob >  \Delta$   \newline \hspace*{30 pt} for any $s^{''} \in$ PASSED} 
\State $\verb+WCET+ := \verb+WCET+ + prob * (|s'.Z_{(CLK, 0)} - s.Z_{(CLK, 0)}|)$
\State $s'.cnt++$
\State \textbf{add} $s'$ \textbf{to} WAIT
\ElsIf {$s'.\ell = s^{''}.\ell \land s^{''}.sts = 1 \land s'.\dot{Z} \neq s^{''}.\dot{Z} \land prob < \Delta$ \newline \hspace*{30 pt} for any $s^{''} \in$ PASSED} 
\State $s'.cnt := 0$
\State $prob := 1 $
\Else 
\State $\verb+WCET+ := \verb+WCET+ + prob * (|s'.Z_{(CLK, 0)} - s.Z_{(CLK, 0)}|)$
\State \textbf{add} $s'$ \textbf{to} WAIT
\EndIf
\EndFor
\EndWhile
\State \textbf{return}  \verb+WCET+
\end{algorithmic}
\end{algorithm}

Algorithm~\ref{alg:EfficientWCETWithAccel} uses an extra clock
\verb+CLK+ to keep track of time delays that can elapse at each state
of the model.  The algorithm uses a number of operations to handle
cycles in the input PTA.  The operation
$ComputeLocationsofCycle()$ is used to compute the set of control
locations of the detected cycle in the form
$(\ell_0,...,\ell_{m-1})$. This is necessary in order to compute the
set of final states when accelerating the execution of the cycle.  The
operation $SynthDelayFormula()$ is used to synthesize a delay
formula for the detected cycle once a fixed point is reached.  Two
acceleration procedures are used, namely $AccelConstCycle()$ which is
used to accelerate cycles with constant delays, and
$AccelPeriodCycle()$ which is used to accelerate cycles with periodic
delays. Each of these acceleration procedures consists of a number of
operations as described in Section~\ref{sec: accelerations}.
Note that in some cases, however, Algorithm~\ref{alg:EfficientWCETWithAccel} may
compute more than one final state when accelerating the execution of a
detected cycle $\pi$, depending mainly on the structure of the cycle.
That is, for each outgoing edge $e_j$ of the cycle's location $\ell$,
where $e_j \not \in E_{\pi}$, the algorithm computes a final state.
So that if there are $k$ control locations of the cycle $\pi$ that
have more than one outgoing edge then the algorithm computes $k$ final
states.  

It is interesting to note also that the algorithm uses the
activity abstraction when searching for a fixed-point of visited
cycles.  The activity abstraction ignores clocks that are inactive at
some point during the exploration.  A clock is active within a cycle
$\pi$ if its value at some location of the cycle may influence the
future evolution of the cycle. This can happen either when the clock
appears in the invariant condition of some location of the cycle, it
is tested in the condition of some of the edges of the cycle, or an
active clock takes its value when moving through an edge of the cycle.
We write $s.\dot{Z}$ to refer to the set of clock constraints
involving active clocks at state $s$.

\begin{theorem}

Algorithm \ref{alg:EfficientWCETWithAccel} computes a sound estimation of WCET of PTAs.

\end{theorem}

To compute the WCET of a cyclic
PTA, Algorithm \ref{alg:EfficientWCETWithAccel} requires 
that each reachable cycle is repeated until the
probability that results from successive executions of the cycle
converges to zero. Since there is actually no end (it is not possible,
theoretically, to reach zero), Algorithm \ref{alg:EfficientWCETWithAccel}
uses an arbitrary stopping point
$\Delta$, which is chosen in a way such that any errors accumulated
across several cycles are minimized and so that zero can be
\emph{effectively} reached. This ensures the sound
estimation of whole automaton WCET. 

\section{Implementation}

In this section we briefly summarise our prototype implementation of the model checking
algorithms given in Section \ref{sec:Algorithm}. 
It is important to note that the goal of
our implementation is to validate the presented algorithms, rather than to
devise an efficient implementation; this will be the subject of our future work.

The prototype implementation has been developed using the opaal tool \cite{opaal} 
which has been  designed to rapidly prototype new model checking algorithms.
The opaal tool is implemented in Python and is a standalone model checking engine.
We use the open source UPPAAL DBM library for the internal symbolic representation of time zones in the algorithms. 

\begin{wrapfigure}{R}{0.5\textwidth} 
\vspace{-20pt}
     \centering
    \includegraphics[width=0.48\textwidth] {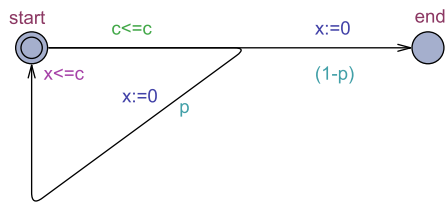}
    \caption{Demonstrating example}
    \label{fig:SymbolicEx}
     \vspace{-20pt}
  \vspace{1pt}
\end{wrapfigure}

We consider here  one example of
cyclic PTA (see Fig. \ref{fig:SymbolicEx}),
but we verify it under four different
settings: (a) when $p = 0.001$ and $c = 1$, 
(b) when $p = 0.001$ and $c = 10^{6}$,
(c) when $p = 0.999$ and $c = 1$,
and (d) when $p = 0.999$ and $c = 10^{6}$.
It is easy to see that the WCET of the automaton
under these four settings will be different,
as the number of times the cycle will be repeated and the time that
can elapse at each iteration will be different. 
For this example, we set $\Delta = 10^{-6}$.
It is easy to see that the cycle in the given automaton has constant delays 
as the active clock of the cycle (clock \verb+x+) is reset each time
the cycle is executed and hence after two iterations 
the search will reach a fixed-point at location \verb+Start+. 
The synthesized delay formula for computing WCET of
the cycle will be $\phi_{\pi} = \sum_{i =1}^{n} (p^{i} * c)$,
where $ n = \dfrac{ln (\Delta)}{ln (p)}$.
The WCET as computed by the algorithm 
for the four cases is as follows: (a) WCET = 1.001,
(b) WCET = 1001001.001,  (c) WCET = 1000, and (d) WCET = $10 ^{9}$.
For cases (a) and (b) the cycle needs to be repeated only two times,
while for cases (c) and (d) the cycle needs to be repeated  about 13808 times.
However, the algorithm collapsed the iterations of the cycle and hence it avoided
the explicit repeated exploration of the cycle. The algorithm
returned an answer for each case almost instantly.
Note that there is no available implementation for the algorithms
presented in \cite{KNPS03,Jovanovi2017}, and hence we were not able
to report any result about their performance in presence of cycles.  
However, the algorithms in \cite{KNPS03,Jovanovi2017} are not optimized to check WCET
of PTAs (specially those which contain cycles that can be repeated very often due to high probabilities.)

\subsection*{Nested Cycles and Intersecting cycles}

 Algorithm \ref{alg:EfficientWCETWithAccel} can handle only cyclic PTAs
that satisfies the flatness assumption, wherein each location 
can be part of at most one cycle.
Hence, nested cycles and intersecting cycles 
(i.e. two or more cycles which have at least
one control location in common) cannot be handled using Algorithm \ref{alg:EfficientWCETWithAccel}.
The presence of such classes of cycles complicates 
the formal verification of expected WCET of PTAs.
In particular, if there is a nested cycle in the automaton, 
then if some inner cycle is detected and then
collapsed, the adjustment of the weights performed (as detailed in section 3)
(along with the addition of the visited states to the PASSED list in the
algorithm) would impact the ability to accurately collapse an ``outer cycle''
with arcs composed in part of the inner cycle. 
Furthermore, the order at which intersecting cycles are executed can affect the outcome
of the WCET analysis, depending on the way the probabilistic choices 
at common control location are resolved. In future work, we aim to extend
the algorithm to handle complex forms of cycles including
nested cycles and intersecting cycles. 

\section{Conclusion and Future Work}

We have described a model checking algorithm which can be applied to
verify expected WCET of probabilistic timed systems with cyclic
behavior. Indeed, the presence of cycles that can be repeated a very
high number of times in the input timed probabilistic model can
degrade the performance of the model checking algorithm. However, we
have shown that it is possible to accelerate the execution of
probabilistic timed cycles without adversely affecting the outcome of
the analysis. In a future work, we aim to reconsider the problem while
allowing non-deterministic choices between edges, where the precise
complexity of the expected WCET problem for cyclic PTAs with
non-determinism is still open.

\bibliographystyle{plain}
\bibliography{references}

\end{document}